\documentclass[aps,prb,superscriptaddress,twocolumn]{revtex4-1}

\usepackage[version=3]{mhchem}
\usepackage{sistyle}
\usepackage{tabularx}
\usepackage[caption=false]{subfig}
\usepackage{hyperref}
\usepackage{xcolor}
\usepackage{natbib}
\hypersetup{pdfhighlight=/N,pdfborder={0 0 0}}

\usepackage{tikz}
\usetikzlibrary{arrows,decorations.pathmorphing,backgrounds,positioning,fit,petri,calc,shapes.geometric,scopes,fit}
\usepackage{pgfplots}
\pgfplotsset{compat = 1.3}
\usepackage{tkz-euclide}
\usetikzlibrary{external}
\usepackage[acronym]{glossaries}
\newacronym[shortplural=f.u.]{fu}{f.u.}{formula unit}

\newacronym{cnt}{CNT}{carbon nanotube}
\newacronym{swcnt}{SWCNT}{single walled carbon nanotube}
\newacronym{mwcnt}{MWCNT}{multi walled carbon nanotube}

\newacronym{cvd}{CVD}{chemical vapor depostion}

\newacronym{fp}{FP}{frozen phonon}
\newacronym{dft}{DFT}{density functional theory}
\newacronym{paw}{PAW}{projector augmented wave}

\newacronym{lda}{LDA}{local density approximation}
\newacronym{gga}{GGA}{generalized gradient approximation}
\newacronym{pbe}{PBE}{Per\-dew--Burke--Ern\-zer\-hof}

\newacronym{dos}{DOS}{density of states}

\makeglossaries

\newcommand\diff{\mathrm d}

\newcommand\externOrTikz[2]{#1}

\begin{document}
\title{Elastic and Piezoresistive Properties of Nickel Carbides from First-Principles}

\newcommand\affHZDR{
\affiliation{
Helmholtz-Zentrum Dresden - Rossendorf,
Institute of Ion Beam Physics and Materials Research,
Bautzner Landstra\ss{}e 400, 01328 Dresden, Germany
}
}
\newcommand\affTUC{
\affiliation{
Institute of Physics, TU Chemnitz,
09107 Chemnitz, Germany}
}
\newcommand\affNanoNet{
\affiliation{
Helmholtz-Zentrum Dresden - Rossendorf,
International Helmholtz Research School for Nanoelectronic Networks (IHRS NanoNet),
Bautzner Landstra\ss{}e 400, 01328 Dresden, Germany}
}
\newcommand\affENAS{
\affiliation{
Fraunhofer Institute for Electronic Nano Systems (ENAS),
Technologie-Campus 3, 09126 Chemnitz, Germany}
}
\newcommand\affDCMS{
\affiliation{
Dresden Center for Computational Materials Science (DCMS),
TU Dresden, 01062 Dresden, Germany}
}
\newcommand\affCFAED{
\affiliation{
Center for Advancing Electronics Dresden (cfaed),
TU Dresden, 01062 Dresden, Germany}
}

\author{Jeffrey~Kelling}
\email{j.kelling@hzdr.de}
\affHZDR\affTUC\affNanoNet
\author{Peter~Zahn}
\affHZDR\affNanoNet\affDCMS
\author{J\"org~Schuster}
\affNanoNet\affDCMS\affENAS\affCFAED
\author{Sibylle~Gemming}
\email{s.gemming@hzdr.de}
\affHZDR\affTUC\affNanoNet\affDCMS\affCFAED

\begin{abstract}
  The nickel--carbon system has received increased attention over the past years
due to the relevance of nickel as a catalyst for carbon nanotube and graphene
growth, where Nickel carbide intermediates may be involved or carbide interface
layers form in the end. Nickel--carbon composite thin films comprising \ce{Ni3C}
are especially interesting in mechanical sensing applications. Due to the
meta-stability of nickel carbides, formation conditions and the coupling
between mechanical and electrical properties are not yet well understood.
Using first-principles electronic structure methods, we calculated the
elastic properties of \ce{Ni3C}, \ce{Ni2C} and \ce{NiC}, as well as changes in
electronic properties under mechanical strain.  We observe that the electronic
density of states around the Fermi level does not change under the considered
strains of up to \SI1\%, which correspond to stresses up to \SI3{GPa}. Relative
changes in conductivity of \ce{Ni3C} range up to maximum values of about
\SI{10}\%.

\end{abstract}

 \maketitle

 \section{Introduction}
  Nickel--carbon compounds and composite thin films containing amorphous carbon
are of high interest for various applications.
Thin films have been investigated for their piezoresistive
properties~\cite{Uhlig2013129} and as low friction solid
lubricants~\cite{schaefer2011bridging}. The meta-stable \ce{Ni3C} has
been frequently observed in such films~\cite{Uhlig201325,He2010,furlan2014_NiCx} and was suggested to
cause piezoresistive behaviour~\cite{Uhlig2013129}. This carbide has been
reported to be hard to distinguish from \texttt{hcp}-nickel, where a
study~\cite{furlan2014_NiCx} suggests that \texttt{hcp}-nickel is only
stable in the presence of carbon and with some carbon content. A meta-study on
this subject can be found in reference~\onlinecite{He2010}. A recent study~\cite{BBMB2016}
confirmed, that \ce{Ni3C} in such films only decomposes at
temperatures well above \SI{250}\degC.

The nickel--carbon system is also of interest for the catalytic production of
\glspl{cnt} and graphene. \gls{cnt}-growth was achieved both using nickel
nanoparticles as a catalyst~\cite{sato2003_cntNi,doi:10.1021/jp1045707} and on
carbon--nickel nanocomposite thin films~\cite{krause2012_swcntCNi}. While
studies suggest, that carbides do not form during \gls{cnt} growth from \ce{Ni}
nanoparticles~\cite{LTBL2006}, \ce{Ni3C} has been observed in nanoparticles
after \gls{cnt} growth by plasma enhanced chemical vapor deposition was
stopped~\cite{Ducati2006}. A more recent study~\cite{ZZHW2008} confirmed, that
\ce{Ni}/\ce{Ni3C} core-shell structures can indeed be produced. In such a setup,
the carbide could act as an advanced
contact material for \gls{cnt} junctions with properties similar to those
demonstrated for \ce{Mo2C}~\cite{CHTF2015}. The advantage would be, that the
\ce{Ni3C}--\gls{cnt} unit can be grown bottom-up.
\ce{Ni3C} does also occur as a parasitic by-product
of carbon nanofiber-growth on nickel foam~\cite{McDonough2009}.

Graphene~\cite{ZTHF2010} and graphene--type interfacial layers~\cite{WHMMG2016}
can be produced by metal-induced crystallization and layer
inversion as well as by epitaxial growth on transition metals, such
as nickel~\cite{GGWS12}. In the latter case, one study~\cite{WBB11} excluded the
occurrence of crystalline \ce{Ni3C} on a polycrystalline \ce{Ni}
surface by XRD measurements. 
Others observed an
interface layer between $\{\mathtt{111}\}$-nickel and graphene with the
stoichiometry \ce{Ni2C} by Auger
spectroscopy~\cite{lahiri2011_ni2cgraphene,jacobson2012_ni2cgraphene}. In both cases,
mechanical details, especially of carbide intermediates require further study.

\begin{figure}[b!h]
 \centering
 \newlength\structHeight
 \newsavebox{\leftStructBox}\savebox{\leftStructBox}{
 \begin{minipage}[b]{\linewidth/3-2em}
  \subfloat[\ce{Ni2C} (058)\label{fig:struct:58}]
 {
  \externOrTikz{\includegraphics{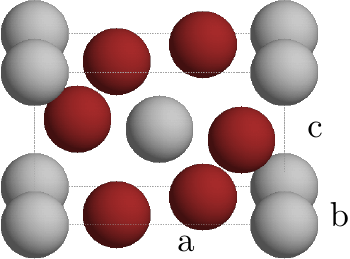}}{%
  \tikzsetnextfilename{ni2c058}%
  \begin{tikzpicture}[every node/.style={outer sep=0em,inner sep=0em}]
   \node[anchor=south west,] at (0,0) {
   \includegraphics[width=\linewidth+2.8em]{bilderRaw/ni2c058.png}};
   \node[anchor=center] at (3.2,1.3) {c};
   \node[anchor=center] at (3.45,.45) {b};
   \node[anchor=west] at (1.8,.15) {a};
  \end{tikzpicture}}
 }
 \\[-.5em]
 \subfloat[\ce{Ni2C} (060)\label{fig:struct:60}]
 {
  \externOrTikz{\includegraphics{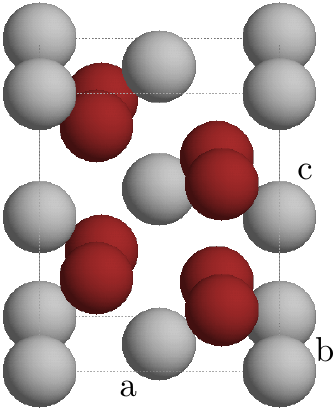}}{%
  \tikzsetnextfilename{ni2c060}%
  \begin{tikzpicture}[every node/.style={outer sep=0em,inner sep=0em}]
   \node[anchor=south west,] at (0,0) {
    \includegraphics[width=\linewidth+2.8em]{bilderRaw/ni2c060.png}};
   \node[anchor=center] at (3.1,2.4) {c};
   \node[anchor=center] at (3.3,.6) {b};
   \node[anchor=center] at (1.3,.2) {a};
  \end{tikzpicture}}
 }
 \end{minipage}
 }
 \usebox\leftStructBox
 \hspace{4em}
 \subfloat[\ce{Ni3C} (167)\label{fig:struct:ni3c}]
 {
  \externOrTikz{\includegraphics{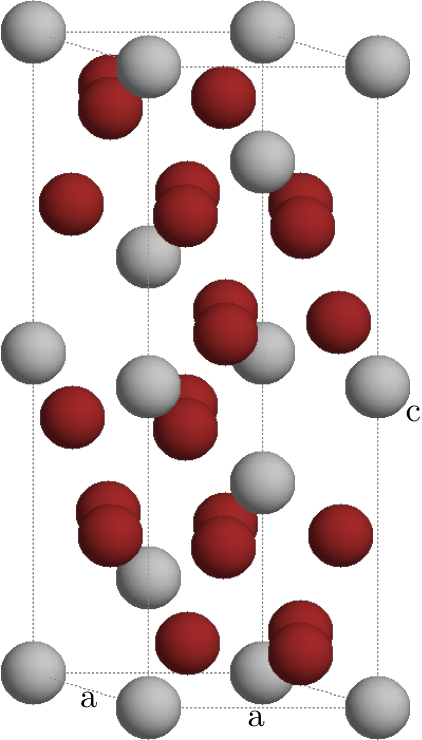}}{%
  \setlength{\structHeight}{218.66455pt}
  \tikzsetnextfilename{ni3c}%
  \begin{tikzpicture}[every node/.style={outer sep=0em,inner sep=0em}]
   \node[anchor=south west] at (0,0) {
    \includegraphics[height=\structHeight-0.5em]{bilderRaw/ni3cHex.png}};
   \node[anchor=center] at (4.2,3.3) {c};
   \node[anchor=center] at (2.6,0.2) {a};
   \node[anchor=center] at (.9,.4) {a};
  \end{tikzpicture}}%
 }
 \caption{(color online) Overview of the crystal structures used in this work for \ce{Ni2C} of
  space groups 058~(\protect\subref{fig:struct:58}, two~\glspl{fu}) and
  060~(\protect\subref{fig:struct:60}, four~\glspl{fu}), both orthorhombic,
  as well as of \ce{Ni3C}~(\protect\subref{fig:struct:ni3c}, space group 167,
  six~\glspl{fu}).
  The latter is displayed in a hexagonal unit
  cell for clarity, where the angle enclosed by the $a$ edges in the basal plane
  is equal to \ang{120}. A primitive rhombohedral cell of only one third the size
  exists which was used for calculations. The grey balls represent carbon and
  the red ones represent nickel. The unit vector $\mathbf{e}_x$ is always
  parallel to $a$ and $\mathbf{e}_z$ is parallel to $c$.
 \label{fig:struct}}
\glsreset{fu}
\end{figure}

The stability of a range of nickel--carbides has been investigated by density
functional calculations~\cite{gibson2010}, yet, neglecting the influence of
elastic deformations which we address here. The study confirmed that, without
externally induced strains, \ce{Ni3C} in space group 167 structure,
figure~\ref{fig:struct}\subref{fig:struct:ni3c}, is the
least unstable carbide and suggests that \ce{Ni2C} is most stable in orthorhombic
structures of space groups 058~(\textit{Pnnm}) and 060~(\textit{Pbcn}), see
figures~\ref{fig:struct}\subref{fig:struct:58} and \subref{fig:struct:60}.

All in all, especially the phase \ce{Ni3C} has potential technical applications in
heterojunctions consisting of nickel and carbon allotropes, including
\glspl{cnt},
acting as electrical circuit elements, for example piezoresistive
sensors~\cite{WagnerSchuster2012}. In these applications, the mechanical and
piezoresistive properties of a potential carbide layer between nickel and the
carbon structure can become relevant when the device is being strained during
operation or when the layer is under constant epitaxial stress which may be
caused the large surface tension of nickel~\cite{EJBAT2005}.

The present work focuses on investigating the elastic properties of the three nickel
carbides \ce{NiC}, \ce{Ni2C} and \ce{Ni3C} in their most stable
crystallographic structures. Ground state properties of the
carbides are compared in section~\ref{s:gs}, the obtained elastic properties are
discussed in section~\ref{s:elast}.
For the experimentally most relevant carbide, \ce{Ni3C}, the influence of strain on
the electronic transport properties is dicussed in section~\ref{s:transport}.

 \section{Computational Methods}
  \subsection{Electronic Structure Calculations}
   \ce{NiC} was calculated in rocksalt (B1)
structure, for \ce{Ni2C} the structures proposed by Gibson et
al.~\cite{gibson2010} were used and for \ce{Ni3C} the rhombohedral (bainite,
space group 167) structure, which was experimentally found by
Nagakura~\cite{JPSJ.12.482} was assumed (see
figure~\ref{fig:struct}\subref{fig:struct:ni3c}).

All results presented here were obtained applying \gls{dft}, in the \gls{gga}
in the \gls{pbe} parametrization~\cite{perdew1996GenGraAppMadSim} as
exchange--correlation functional, which is known to give good results for
bulk mechanical properties when comparing to experiments~\cite{kurth1999_PBEelast}.
The plane-wave im\-ple\-men\-ta\-tion in the ABINIT
pack\-age~\cite{gonze2009abinit1,gonze2002abinit2,gonze2005abinit0} was used,
employing the \gls{paw} method~\cite{torrent2008abinitPAW}. The \gls{paw} atomic
data sets treat $3\mathrm d^8 4\mathrm s^2$ and $2\mathrm s^2 2\mathrm p^2$ as
valence for nickel~\cite{PAW_ni} and carbon~\cite{PAW_cHard}, respectively.

For numerical accuracy, the plane-wave cutoff was converged to
$E_\mathrm{cut}\sim\SI{980}{eV}$ ($=\SI{36}{Ha}$). Only, for calculations of
\texttt{fcc}-nickel, the cutoff was set to about $\SI{1360}{eV}$ ($\SI{50}{Ha}$)
in order to reach a convergence of total energy below about $\SI{2.7}{meV}$
(\SI{1e-4}{Ha}) per atom.  At this point energy differences under strain are
converged to below about $\SI{0.27}{\mu{}eV}$ (\SI{1e-8}{Ha}), which is far more
accurate than required for structural relaxation and the calculation of elastic
properties. The stronger total energy criterion was chosen with regard to
calculating formation enthalpies.

When calculating ground state properties of carbides the Brillouin zone was
sampled with a Monkhorst--Pack grid of $12\times12\times12$ $k$-points.
Thermal smearing of Fermi-Dirac-type~\cite{ABINIToccopt} was fixed to about
$\SI{27}{meV}$ ($\SI{1e-3}{Ha}$). Since the
unit cells of nickel and diamond are smaller, denser grids of
$32\times32\times32$ and $16\times16\times16$ $k$-points, respectively, were required in order
to get comparable sampling accuracy.

The ground state formation energies per \gls{fu} for the carbides were
calculated according to
\begin{equation}
 \Delta E_f = E_{\ce{Ni_xC_y}} -
 x\,E_{\text{\texttt{fcc}-\ce{Ni}}} - y\,(E_\mathrm{diamond} -
 \SI{25}{meV})\qquad,
 \label{eq:efrom}
\end{equation}
where $E_i$ is the total energy of compound $i$.
Diamond was calculated as carbon reference structure instead of graphite
because the employed method is not capable of correctly calculating van-der-Vaals
interactions. An empiric correction of $\Delta E_{\ce{C}} = \SI{25}{meV}$ per
carbon atom, also used in reference~\onlinecite{gibson2010}, was applied to obtain formation
energies with respect to graphite.

  \subsection{Frozen Phonon Calculations}
   Within the linear regime, elastic properties can be described by the elastic
tensor $\hat C$, which gives the stress response $\hat\sigma$ of a material
proportional to a deformation $\hat e$:
\begin{align}
 \sigma_i &= \sum\limits_j c_{ij}\cdot e_j
 \label{eq:sig_ce}
 \intertext{Here, Voigt's notation is used to write the stress and deformation tensors as
six-vectors ($11\to1$; $22\to2$; $33\to3$; $23\to4$; $13\to5$; $12\to6$), with
entries corresponding to three axial strains ($1-3$) and shear strains ($4-6$).
In this
way the elastic tensor can be written as a $6\times6$ matrix from which all
elastic properties can be derived. The bulk modulus is given by: }
B &= \frac13 \left(\left\langle c_{11}\right\rangle + 2\left\langle
c_{12}\right\rangle\right)
 \label{eq:bulkmod}
\end{align}
where $\left\langle c_{11}\right\rangle$ denotes an average over the
 diagonal axial strain entries and $\left\langle c_{12}\right\rangle$ an average
 over the off--diagonal axial strain entries.

The entries of the elastic tensor were calculated using the \gls{fp} method, where the
stress-response was derived from ground state calculations of the deformed
primitive cell. A more detailed explanation can be found
in~\cite{Wagner20086232}. The six primitive deformations were applied separately
with magnitudes ranging up to \SI1\%. All elastic constants were then determined
using equation~\eqref{eq:sig_ce}. The diagonal entries of the tensor can also be
determined from the total energies of the same calculations:
\begin{equation}
 E_\mathrm{\delta} = E_0 + \frac{V_0}2\,\sum\limits_{i} c_{ii} \cdot e_i^2
 \label{eq:e_ce}
\end{equation}
where $E_0$ and $V_0$ are the total energy and volume of the unstrained cell.
The calculated tensors were checked for consistency by comparing the results
of equations~\eqref{eq:sig_ce} and~\eqref{eq:e_ce}. The calculation parameters were
converged until the difference between the diagonal tensor elements from both
equations was less than~\SI2{GPa}. This criterion called for using a
$48\times48\times48$ $k$-point grid for the deformed cell of \ce{NiC}, for the
other materials it was met by using the aforementioned simulation parameters.

If the material's unit cell exhibits internal degrees of freedom, performing a ground
state calculation of the deformed cell without relaxation of the ion positions
yields entries of the so-called
\emph{clamped--ion} elastic tensor $\hat C^c$. To obtain the more physical
\emph{relaxed--ion} elastic tensor $\hat C^r$, the internal atomic coordinates were
relaxed using the Broyden--Fletcher--Goldfarb--Shanno algorithm as implemented
in ABINIT until all forces were below \SI{5e-4}{eV/\angstrom}.

  \subsection{Electronic Transport}
   Electronic transport was calculated assuming constant relaxation time $\tau$
within the Boltzmann formalism where the conductivity tensor at zero temperature
is given as:~\cite{Ziman1972}
\begin{equation}
 \varsigma_{ij} = \tau \frac{e^2}{(2\pi)^3\hbar} \sum\limits_n
 \int\limits_{\varepsilon^n(\mathbf k) = E_\mathrm{Fermi}} \hspace{-2em}\diff S
 \frac {v^n_i(\mathbf k) v^n_j(\mathbf k)} {\left| \mathbf v^n(\mathbf
 k)\right|}\label{eq:boltzmannT}
\end{equation}
\[
 \quad\text{with}\quad\mathbf v^n(\mathbf k) = \frac 1\hbar
 \nabla_\mathbf{k} \varepsilon^n(\mathbf k)
\]
where $\varepsilon^n$ is the eigenenergy of the $n$th band and
$\mathbf{v}^n(\mathbf{k})$ the vector of the corresponding group velocity. $e$
denotes the electron charge and $i,j$ denote cartesian vector components.

Off-diagonal elements of $\varsigma_{ij}$ are zero by symmetry.  For the
relaxation time $\tau$ no specific value is assumed, though it might be
anisotropic ($\tau_{zz}\neq\tau_{xx}=\tau_{yy}$) in the case of $\ce{Ni3C}$ due
to its rhombohedral structure~\cite{YHMZ2011}. The integrals on the right-hand
side of equation~\eqref{eq:boltzmannT} reflect the anisotropy of the
bandstructure of the unperturbed, but eventually strained, systems at the Fermi
level. Assuming that $\tau$ remains constant under strain in the linear regime,
since no new scattering centers are created,
predictions can be made about the change of conductivity under strain.

For strained cells, the bandstructure with relaxed ion positions was used as
basis for these calculations.

 \section{Results and Discussion}
  \subsection{Ground State Results\label{s:gs}}
   The lattice parameters of the investigated materials are available in literature,
some even from experiments. The lattice constant calculated for
\texttt{fcc}-nickel in the present work ($a_\ce{Ni}=\SI{3.524}\angstrom$) agrees
very well with values found in
literature~\cite{PhysRev.10.661,gibson2010}. The obtained lattice parameter for
diamond ($a_\mathrm{diamond}=\SI{3.577}\angstrom$) is only slightly larger than
the experimental value of
$\SI{3.567}\angstrom$~\cite{PhysRevB.49.9341}.
Lattice parameters obtained for the carbides as well as formation enthalpies
will be given for comparison, the latter with respect to \texttt{fcc}--\ce{Ni}
and graphite.

\paragraph{\ce{NiC}} Assuming rocksalt structure,
the lattice parameter $a_\ce{NiC} = \SI{4.073}\angstrom$ was obtained, which is
in good agreement with ref.~\cite{gibson2010} (\SI{4.077}\angstrom) and other numerical studies cited
therein. The calculated formation enthalpy of $\Delta E_{f,\ce{NiC}} =
\SI{49.7}{kcal/mol}$ of~\gls{fu} also agrees with ref.~\cite{gibson2010} (\SI{48.6}{kcal/mol}).

\begin{table}[ht]
 \caption{Lattice parameters, formation enthalpies and total energies for the
  two considered orthorhombic structures of \ce{Ni2C}. Length values
  are given in~\SI{}{\angstrom}, formation enthalpies in \SI{}{kcal/mol} of
  \gls{fu}
  Values from ref.~\cite{gibson2010} are given in parentheses.
 \label{tab:ni2c}}
 \centering
 \begin{tabular}{lrr}
  \hline
  &\ce{Ni2C} (058) &\ce{Ni2C} (060) \\\hline
  $a$ & \num{4.72}(\num{4.72}) & \num{4.19}(\num{4.19}) \\
  $b$ & \num{4.19}(\num{4.17}) & \num{5.51}(\num{5.51}) \\
  $c$ & \num{2.93}(\num{2.92}) & \num{4.94}(\num{4.94}) \\
  $\Delta E_{f}$ & \num{12.2}(\num{7.9}) & \num{12.0}(\num{7.9}) \\
  \hline
 \end{tabular}
\end{table}

\paragraph{\ce{Ni2C}} The calculated values for the two investigated structures
are summarized in table~\ref{tab:ni2c}. The
lattice parameters are in excellent agreement with ref.~\cite{gibson2010}. Only, the
formation enthalpies stated therein disagree with the present results (see
table~\ref{tab:ni2c}, values in parentheses). However,
there is agreement on the prediction that both structures are essentially
degenerate, with the variant of space group 060 being less than $\SI{5}{meV}$ lower in
total energy.

\paragraph{\ce{Ni3C}} The obtained lattice parameters $a=\SI{4.60}{\angstrom}$
and $c=\SI{13.00}\angstrom$ are in good agreement with ref.~\cite{gibson2010}
($a=\SI{4.49}{\angstrom}$, $c=\SI{13.02}\angstrom$) and electron diffraction
measurements~\cite{JPSJ.13.1005} ($a=\SI{4.553}{\angstrom}$,
$c=\SI{12.92}\angstrom$). A formation enthalpy of $\Delta E_{f,\ce{Ni3C}} =
\SI{6.3}{kcal/mol}$ was obtained, which is identical to the value reported in
ref~\cite{gibson2010} and reflects the thermal decomposition observed in~\cite{CCS2014}.

\begin{figure}
 \centering
 \externOrTikz{%
 \includegraphics[width=8.5cm]{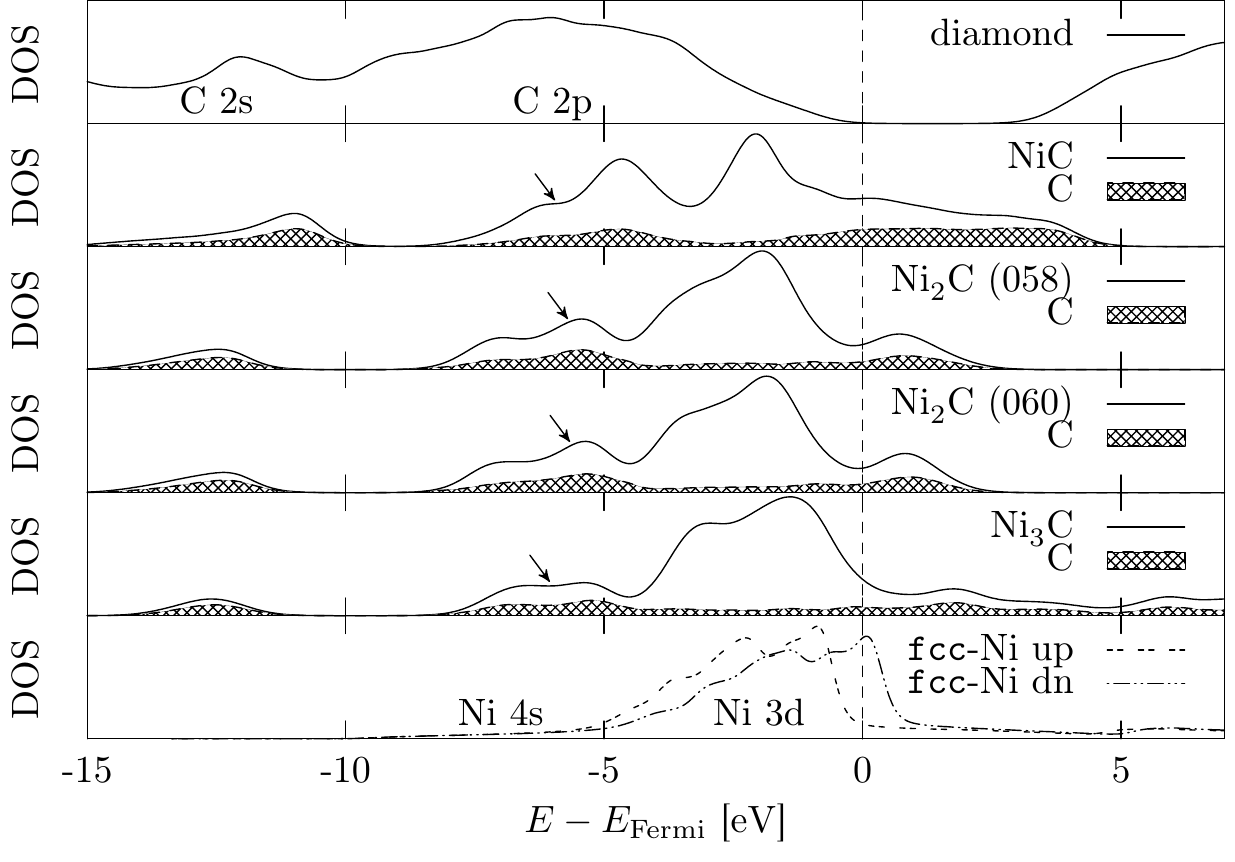}}{%
 \includegraphics[width=8.5cm]{dos.pdf}}
 \caption{
  Total and projected \gls{dos} of carbides compared to those of the reference phases. For display,
  Gaussian smearing of about $\SI{270}{meV}$ ($\SI{1e-2}{Ha}$) was applied.
  Atomic orbitals are indicated as obtained from
  calculations of projected \gls{dos}, filled areas under curves indicate the
  fraction of \gls{dos} attributed to \ce C sites.
  The arrows indicate the bonding band with
 \ce C 2p and hybrid \ce{Ni} states.
 \label{fig:dos}}
\end{figure}

For the relaxed, strain-free geometries, all carbides of nickel investigated
here are meta-stable at $T=\SI0K$, non-magnetic and metallic as observed
previously~\cite{gibson2010}. \ce{Ni2C} exhibits a very low \gls{dos} around the
Fermi energy. The \gls{dos} for the investigated carbides and
the reference phases are plotted in figure~\ref{fig:dos}. 

In all carbides the \ce C 2s band is located below the conduction band.
It is shifted to lower energies (shifted left in figure~\ref{fig:dos}) for
\ce{Ni2C} and \ce{Ni3C} in comparison to the carbide with higher carbon content,
\ce{NiC}, indicating a deeper potential well for electrons is provided by the
carbon atoms. They are also more strongly negatively charged than in
\ce{NiC}.
The \ce{Ni} 3d states are located below
the Fermi energy above about $\SI{-5}{eV}$ for \ce{Ni2C} and \ce{Ni3C}, for
\ce{NiC} they are spread over a broader energy range, starting
at around $\SI{-6}{eV}$. The \ce{Ni} 3d orbitals do contribute to the \gls{dos} at the
Fermi level, but much less than in \texttt{fcc}-nickel, where the 3d-\gls{dos} of the
minority spin peaks at the Fermi level. \ce{Ni} 4s and \ce{C} 2p states also
contribute to the \gls{dos} at the Fermi level. The part of the conduction band
below about $\SI{-5}{eV}$ is predominately composed of \ce C 2p states hybridizing
with \ce{Ni} states, see arrows in figure~\ref{fig:dos}.

  \subsection{Elastic Constants\label{s:elast}}
   As a reference, the elastic tensors of \texttt{fcc}-\ce{Ni} and diamond were
calculated and the non-zero, not symmetrically equivalent elements are provided in
table~\ref{tab:elast}. The calculated bulk modulus for diamond is identical to
earlier theoretical works~\cite{PhysRevB.32.7988} and also the tensor components agree
with earlier literature data~\cite{PhysRevLett.84.5160}. The bulk modulus for \ce{Ni} is
within about \SI{10}{GPa} of experimental results~\cite{RSAJ2001}.
This deviation is predominantly
attributed to the approximations involved in \gls{dft}. The following
predictions for the elastic properties of nickel carbides can be expected to
have about the same accuracy.

All carbides exhibit a larger bulk modulus than Nickel and a much lower one than
diamond, as apparent from the last column of table~\ref{tab:elast}. Being the
carbide with the highest carbon content, \ce{NiC} shows the
largest bulk modulus of the carbides. Evidently, the bulk modulus increases with
increasing carbon content, that is, the substances become harder.
Table~\ref{tab:elast} lists all calculated non-zero and not symmetrically
equivalent elastic constants. The carbides \ce{Ni2C} and \ce{Ni3C} exhibit less
symmetric unit cells, resulting in more independent entries in the elastic
tensor.

\paragraph{\ce{Ni2C}} Both investigated hypothetical forms of \ce{Ni2C} are
predicted to be equally hard and even show quite similar anisotropies, probably
due to the fact that both are orthorhombic. The elastic properties of the sample
should not depend on the relative prevalence of these two phases. Still, judging
by the elastic tensors, deforming one cell into the equilibrium shape of the
other and allowing the atoms to rearrange into the other structure by relaxation
requires overcoming a large potential barrier. Thus, even under stress, both
structures can be expected to coexist in one sample.

\paragraph{\ce{Ni3C}} Judging by the obtained bulk moduli, a macroscopically
isotropic polycrystalline sample of \ce{Ni3C} is predicted to be about as hard
as the less stable \ce{Ni2C}.
Even for the most extreme simulated deformations of \SI1\%
stresses were found to be still in the linear regime. Using the calculated value for
$c_{11}^r$, a compression in $\mathbf e_1$--direction of this magnitude corresponds to
applying a pressure of about $\SI{2.7}{GPa}$, which by far exceeds pressures
achievable in most experiments.

Investigating \gls{dos} and band structure of deformed cells, no qualitative
difference with respect to that obtained for the equilibrium geometry was found.
For purely axial strains and compressions (i.e.~$\mathbf e_1$, $\mathbf e_2$ and $\mathbf
e_3$, see figure~\ref{fig:bands}\subref{fig:bands:zz}) bands move slightly closer to
the Fermi-level under strain and further away under compression. This can be
attributed to changing overlaps between atomic orbitals. This difference is
marginal close to and at the Fermi level, the region most relevant to transport
properties. For pure shear deformations no
significant changes are observed, see figure~\ref{fig:bands}\subref{fig:bands:yz}.
Band structures shown in figure~\ref{fig:bands} for deformed cells use relaxed ion
positions. Clamped cells show qualitatively identical changes, only the shift of
bands for strains and compressions is larger.

\begin{table*}[t]
 \caption{Calculated elastic constants, $c_{ij}$ and bulk modulus $B$, for the
  considered carbide phases and the reference phases diamond and
  \texttt{fcc}-Nickel in \SI{}{GPa}. For phases with atomic degrees of freedom
  on the unit cell (\ce{Ni2C}, \ce{Ni3C}) both clamped-ion and relaxed-ion results
  are listed. All omitted entries are given by or are zero by symmetry. For
  space group 167 (\ce{Ni3C}) the following relation holds: $c_{56} = c_{14} =
  -c_{24}$.
 \label{tab:elast}}
 \centering
 \begin{tabular}{lrrrrrrrrrrrr}
  \hline
                           & $c_{11}$   & $c_{22}$ & $c_{33}$ & $c_{12}$  & $c_{13}$ & $c_{23}$ & $c_{44}$  & $c_{55}$ & $c_{66}$ & $c_{14}$ & $B$ \\\hline
  diamond                  & \num{1049} &          &          & \num{129} &          &          & \num{564} &          &          &          & \num{435} \\
  \ce{NiC}                 & \num{296}  &          &          & \num{231} &          &          & \num{50} & & & & \num{256} \\
  \ce{Ni2C} (058), clamped & 316        & 262      & 378      & 218       & 175      & 193      & 116      & 91 &  145 & & 236 \\
  \ce{Ni2C} (058), relaxed & 307        & 234      & 344      & 203       & 160      & 163      & 88 & 87 &  145 & & 215 \\
  \ce{Ni2C} (060), clamped & 279        & 343      & 359      & 213       & 203      & 171      & 90 & 125 &  135 & & 239 \\
  \ce{Ni2C} (060), relaxed & 251        & 333      & 335      & 205       & 186      & 163      & 78 & 91 &  113 & & 225 \\
  \ce{Ni3C} clamped        & 321        &          & 309      & 176       & 184      &          & 116 && 72 & -11 & 227\\
  \ce{Ni3C} relaxed        & 272        &          & 276      & 157       & 150      &          & 91 & & 57 & -22 & 219 \\
  \texttt{fcc}-\ce{Ni}     & \num{266}  &          &          & \num{156} &          &          & \num{129} & & & & \num{192} \\
 \hline
 \end{tabular}
\end{table*}

\def\plotFilename{data/bandsEE0.dat}
\begin{figure}[h!t]
 \centering
 \subfloat[equilibrium\label{fig:bands:ee0}]
 {
  \externOrTikz{\includegraphics{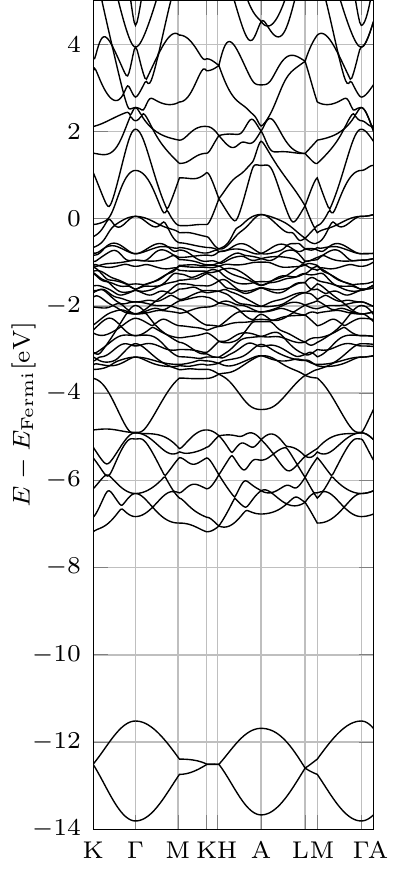}}{%
  \tikzsetnextfilename{bands}%
  \begin{tikzpicture}[
   ]
  \begin{axis}[width=\linewidth/2,height=10cm,
    xtick = {0.,
5.77292534162707e-01,
1.15670548430259e+00,
1.55086523945390e+00,
1.70086523945390e+00,
2.29485777361661e+00,
2.89576537425377e+00,
3.06246537425377e+00,
3.66337297489093e+00,
3.83007297489093e+00},
xticklabels = {K,
$\Gamma$,
M,
{K~},
{~~H},
A,
{L~~},
{~M},
{$\Gamma$~},
{~A}},
grid=both,
font=\scriptsize,
every plot/.style={no marks,black},
ymax=5, ymin=-14, xmin=0, xmax=3.83007297489093e+00,
ylabel={$E-E_\mathrm{Fermi} [\SI{}{eV}]$},
ylabel shift=-.8em,
   ]
\addplot[] table[x index=3, y index=4] {\plotFilename};
\addplot[] table[x index=3, y index=5] {\plotFilename};
\addplot[] table[x index=3, y index=6] {\plotFilename};
\addplot[] table[x index=3, y index=7] {\plotFilename};
\addplot[] table[x index=3, y index=8] {\plotFilename};
\addplot[] table[x index=3, y index=9] {\plotFilename};
\addplot[] table[x index=3, y index=10] {\plotFilename};
\addplot[] table[x index=3, y index=11] {\plotFilename};
\addplot[] table[x index=3, y index=12] {\plotFilename};
\addplot[] table[x index=3, y index=13] {\plotFilename};
\addplot[] table[x index=3, y index=14] {\plotFilename};
\addplot[] table[x index=3, y index=15] {\plotFilename};
\addplot[] table[x index=3, y index=16] {\plotFilename};
\addplot[] table[x index=3, y index=17] {\plotFilename};
\addplot[] table[x index=3, y index=18] {\plotFilename};
\addplot[] table[x index=3, y index=19] {\plotFilename};
\addplot[] table[x index=3, y index=20] {\plotFilename};
\addplot[] table[x index=3, y index=21] {\plotFilename};
\addplot[] table[x index=3, y index=22] {\plotFilename};
\addplot[] table[x index=3, y index=23] {\plotFilename};
\addplot[] table[x index=3, y index=24] {\plotFilename};
\addplot[] table[x index=3, y index=25] {\plotFilename};
\addplot[] table[x index=3, y index=26] {\plotFilename};
\addplot[] table[x index=3, y index=27] {\plotFilename};
\addplot[] table[x index=3, y index=28] {\plotFilename};
\addplot[] table[x index=3, y index=29] {\plotFilename};
\addplot[] table[x index=3, y index=30] {\plotFilename};
\addplot[] table[x index=3, y index=31] {\plotFilename};
\addplot[] table[x index=3, y index=32] {\plotFilename};
\addplot[] table[x index=3, y index=33] {\plotFilename};
\addplot[] table[x index=3, y index=34] {\plotFilename};
\addplot[] table[x index=3, y index=35] {\plotFilename};
\addplot[] table[x index=3, y index=36] {\plotFilename};
\addplot[] table[x index=3, y index=37] {\plotFilename};
\addplot[] table[x index=3, y index=38] {\plotFilename};
\addplot[] table[x index=3, y index=39] {\plotFilename};
\addplot[] table[x index=3, y index=40] {\plotFilename};
\addplot[] table[x index=3, y index=41] {\plotFilename};
\addplot[] table[x index=3, y index=42] {\plotFilename};
\addplot[] table[x index=3, y index=43] {\plotFilename};
\addplot[] table[x index=3, y index=44] {\plotFilename};
\addplot[] table[x index=3, y index=45] {\plotFilename};
\addplot[] table[x index=3, y index=46] {\plotFilename};
  \end{axis}
 \end{tikzpicture}}}
  \begin{minipage}[b]{\linewidth/2-.0ex}
   \externOrTikz{}{%
   \tikzstyle{ls2}=[red, densely dashdotted]
   \tikzstyle{ls3}=[blue, densely dashed]
   \def\plotFilename{data/bandsPapEE0.dat}
   \def\rightPlotHeight{5.15cm}
  }
  \subfloat[$e_3$ (strain) \label{fig:bands:zz}]
  {
  \externOrTikz{\includegraphics{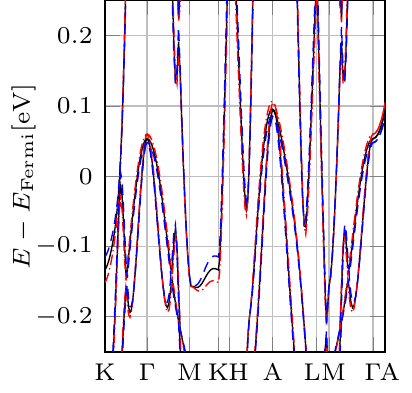}}{%
   \tikzsetnextfilename{bandsZZ}%
   \begin{tikzpicture}[
   ]
  \begin{axis}[width=\linewidth,height=\rightPlotHeight,
    xtick = {0.,
5.77292534162707e-01,
1.15670548430259e+00,
1.55086523945390e+00,
1.70086523945390e+00,
2.29485777361661e+00,
2.89576537425377e+00,
3.06246537425377e+00,
3.66337297489093e+00,
3.83007297489093e+00},
xticklabels = {K,
$\Gamma$,
M,
{K~},
{~~H},
A,
{L~~},
{~M},
{$\Gamma$~},
{~A}},
grid=both,
font=\scriptsize,
every plot/.style={no marks,},
ymax=.25, ymin=-.25, xmin=0, xmax=3.83007297489093e+00,
ylabel={$E-E_\mathrm{Fermi} [\SI{}{eV}]$},
ylabel shift=-.7em,
max space between ticks=20cm,
   ]
   \def\plotFilename{data/bandsPapEE0.dat}
\addplot[,black,] table[x index=3, y index=4] {\plotFilename};
\addplot[,black,] table[x index=3, y index=5] {\plotFilename};
\addplot[,black,] table[x index=3, y index=6] {\plotFilename};
\addplot[,black,] table[x index=3, y index=7] {\plotFilename};
   \def\plotFilename{data/bandsPapZZ010.dat}
\addplot[,ls2] table[x index=3, y index=4] {\plotFilename};
\addplot[,ls2] table[x index=3, y index=5] {\plotFilename};
\addplot[,ls2] table[x index=3, y index=6] {\plotFilename};
\addplot[,ls2] table[x index=3, y index=7] {\plotFilename};
   \def\plotFilename{data/bandsPapZZ-010.dat}
\addplot[,ls3] table[x index=3, y index=4] {\plotFilename};
\addplot[,ls3] table[x index=3, y index=5] {\plotFilename};
\addplot[,ls3] table[x index=3, y index=6] {\plotFilename};
\addplot[,ls3] table[x index=3, y index=7] {\plotFilename};
  \end{axis}
\end{tikzpicture}}
}\\
  \subfloat[$e_4$ (shear) \label{fig:bands:yz}]
  {
  \externOrTikz{\includegraphics{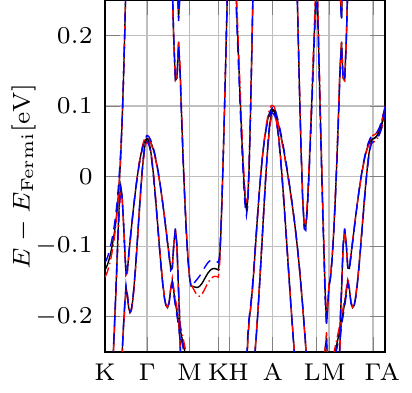}}{%
   \tikzsetnextfilename{bandsYZ}%
   \begin{tikzpicture}[
   ]
  \begin{axis}[width=\linewidth,height=\rightPlotHeight,
    xtick = {0.,
5.77292534162707e-01,
1.15670548430259e+00,
1.55086523945390e+00,
1.70086523945390e+00,
2.29485777361661e+00,
2.89576537425377e+00,
3.06246537425377e+00,
3.66337297489093e+00,
3.83007297489093e+00},
xticklabels = {K,
$\Gamma$,
M,
{K~},
{~~H},
A,
{L~~},
{~M},
{$\Gamma$~},
{~A}},
grid=both,
font=\scriptsize,
every plot/.style={no marks,},
ymax=.25, ymin=-.25, xmin=0, xmax=3.83007297489093e+00,
ylabel={$E-E_\mathrm{Fermi} [\SI{}{eV}]$},
ylabel shift=-.7em,
max space between ticks=20cm,
   ]
   \def\plotFilename{data/bandsPapEE0.dat}
\addplot[,black,] table[x index=3, y index=4] {\plotFilename};
\addplot[,black,] table[x index=3, y index=5] {\plotFilename};
\addplot[,black,] table[x index=3, y index=6] {\plotFilename};
\addplot[,black,] table[x index=3, y index=7] {\plotFilename};
   \def\plotFilename{data/bandsPapYZ010.dat}
\addplot[,ls2] table[x index=3, y index=4] {\plotFilename};
\addplot[,ls2] table[x index=3, y index=5] {\plotFilename};
\addplot[,ls2] table[x index=3, y index=6] {\plotFilename};
\addplot[,ls2] table[x index=3, y index=7] {\plotFilename};
   \def\plotFilename{data/bandsPapYZ-010.dat}
\addplot[,ls3] table[x index=3, y index=4] {\plotFilename};
\addplot[,ls3] table[x index=3, y index=5] {\plotFilename};
\addplot[,ls3] table[x index=3, y index=6] {\plotFilename};
\addplot[,ls3] table[x index=3, y index=7] {\plotFilename};
  \end{axis}
\end{tikzpicture}}
}
 \end{minipage}
 \caption{(color online)
  Band structure of equilibrium \ce{Ni3C}
  for a large energy range~\protect\subref{fig:bands:ee0}
 and for a small interval around $E_\mathrm{Fermi}$ in
  comparison with \SI1\% uniaxially distorted cells (\protect\subref{fig:bands:zz}
  and~\protect\subref{fig:bands:yz}). The displayed deformation directions are
  the ones
  corresponding to the largest strain~\protect\subref{fig:bands:zz} and
  shear~\protect\subref{fig:bands:yz} components of the relaxed elastic tensor.
  Bands of positively deformed cells are plotted in red dash-dotted lines, negative
  deformations in blue dashed lines.
 \label{fig:bands}}
\end{figure}

  \subsection{Electronic Transport under Strain\label{s:transport}}
   A closer analysis of the \ce{Ni3C} band structure yields a density of states of
\SI{2.83}{eV^{-1}} per \gls{fu} and anisotropic averaged Fermi velocities of $v_x = v_y =
\SI{.90e6}{m/s}$ and $v_z=\SI{1.10e6}{m/s}$. This results in an
in-plane/out-of-plane conductivity anisotropy of about $\num{.67}$.

The respective effect of axial strains in $x$ and $z$ direction ($\mathbf{e_1}$
and $\mathbf{e_3}$) on electronic transport was
investigated. For strains up to $\pm1\%$ the
\gls{dos} remains unaffected within the precision of the calculation.
Table~\ref{tab:transport} summarizes relative changes under strain in both
conductance and Fermi velocities. The strongest changes in conductance can be
observed for strains along $\mathbf{e}_3$, where the parallel conductance in $z$
direction changes by $10\%$, the anisotropy increases slightly. For small
strains along $\mathbf{e}_1$ the in-plane isotropy ($xx,yy$) is unaffected,
while the in-plane/out-of-plane anisotropy changes slightly.

Computational and analytical investigations of the piezoresistivity of semi-conducting
\glspl{cnt} suggest an increase in resistance of well above \SI{50}\% under
longitudinal strains of about \SI{.3}\%~\cite{DMDV2015,WagnerSchuster2016}. To
exert the required stress on a \ce{Ni3C}-contacted \gls{cnt}, the contact
material would be strained by about \SI{1}\%, given the ratio of the established
Young's modulus of \glspl{cnt} of about \SI{1000}{GPa}~\cite{WHWHR2008} and the
elastic coefficients we obtained for axial strains in \ce{Ni3C}. This would
cause a change in conductivity of 3 to \SI{10}\% in the contact material, which
is significantly smaller than the effect observed in the \gls{cnt}. Thus, the
piezoresistive properties of a device with \ce{Ni}-contacted \glspl{cnt} as
functional structure are dominated by the electronic response of the
\glspl{cnt} to mechanical deformations.

For isotropic compression of a polycrystalline sample, linear combination of the
axial effects suggests a reduction of conductivity. This is mostly a result of the
relatively large decrease in conductivity by axial compression in $z$ direction.

The piezoresistive effect observed by Uhlig et al. in \ce{Ni3C}--containing
nickel--carbon thin films~\cite{Uhlig2013129,Uhlig201325} under hydrostatic pressure
is opposite to this prediction for bulk \ce{Ni3C} based on our
calculations. Thus our study excludes the possibility, that these observations
are dominated by bulk effects in \ce{Ni3C} grains. One may speculate that they
emerge from effects in nickel grains, since nickel itself is known to show piezoresistive
effects~\cite{PhysRev.94.61,Klokholm1973}. Effects at interfaces in the nickel--carbon
mixture may also play a role.

\begin{table}[b!h]
 \caption{Relative changes in transport properties of \ce{Ni3C} under strain in
  $\mathbf{e}_1$ ($x$) and $\mathbf{e}_3$ ($z$) direction, respectively. Within the accuracy
  of the calculations, transport coefficients in $x$ and $y$ directions are
  affected equally by the considered strain values.
 \label{tab:transport}}
 \centering
 \begin{tabular}{rrrrr}
  \hline
  \multicolumn{1}{l}{strain} & $\Delta\varsigma_{xx}/\varsigma_{xx}$ &
 $\Delta\varsigma_{zz}/\varsigma_{zz}$ & $\Delta v_{x}/v_{x}$ & $\Delta
 v_{z}/v_{z}$ \\\hline
 $e_1 = +1\%$ &
  $-3\%$ & $0$ & $-2\%$ & $-\num{.4}\%$
 \\
 $e_1 = -1\%$ &
  $+3\%$ & $0$ & $+2\%$ & $+\num{.6}\%$ 
 \\
 $e_3 = +1\%$ &
  $+4\%$ & $+10\%$ & $+1\%$ & $+4\%$ \\\hline
 \end{tabular}
\end{table}

 \section{Conclusions}
  The complete sets of elastic constants of nickel carbides have been calculated in
a way that they can be expected to be within \SI{10}{GPa} of experimental
values. The electronic structure and electronic transport properties
of bulk \ce{Ni3C} under stress have been investigated.  Assuming a constant
relaxation time $\tau$, changes in conductivity not exceeding
about $\SI{4}\%$ in-plane and about $\SI{10}\%$ out-of-plane for stresses below
$\SI{3}{GPa}$ are predicted. As a contact material in sensing applications these
changes are of minor significance. These results also show, that \ce{Ni3C} does
not contribute significantly to the piezoresistive effects observed in nickel--carbon thin
films by Uhlig and coworkers~\cite{Uhlig2013129}.

For the formation enthalpy of both \ce{Ni2C} variants, the absolute values
obtained here differ quantitatively from those by Gibson et
al.~\cite{gibson2010}, but qualitatively both studies agree on the relative
ordering with respect to the other carbide phases. On the enthalpy difference
between the ground state \ce{Ni2C} variants the agreement
is excellent. The studies also agree on the formation enthalpies of the other
carbides.

 \section*{Acknowledgments}
  We thank R. Wenisch for fruitful discussions.
This work has been partially financed by the Initiative and Networking
Fund of the German Helmholtz Association via the Helmholtz International
Research School NanoNet \mbox{(VH-KO-606)} and the W2/W3 Programm f\"ur exzellente
Wissenschaftlerinnen \mbox{(W2/W3-026)}. We gratefully acknowledge partial
funding by the DFG via Research Unit FOR1713 (SMINT) and the Center for
Advancing Electronics Dresden (cfaed). We thank the HZDR computing center for
provided computational resources.

 \bibliography{article}
 \bibliographystyle{unsrtnat}
\end{document}